\documentclass{article}

\usepackage{amsthm}
\usepackage{graphicx}
\usepackage{float}
\usepackage{color}
\usepackage{amsmath}
\usepackage{amssymb}
\usepackage{multirow}
\usepackage{multicol}
\usepackage{authblk}
\usepackage{array}
\usepackage{epstopdf}
\usepackage{mathtools}

\newcommand{\E}{\mathbb{E}}

\author[1]{Ryan Murray}
\author[2]{Glenn Young}
\affil[1]{Department of Mathematics, North Carolina State University, Raleigh, NC 27695}
\affil[2]{Department of Mathematics, Kennesaw State University, Marietta, GA 30060}
\date{}                     %% if you don't need date to appear
\begin{document}

\title{Neutral competition in a deterministically changing environment: revisiting continuum approaches}
\date{\today}

\maketitle

\section*{Abstract}

Environmental variation can play an important  role in ecological competition by influencing the relative advantage between competing species. Here, we consider such effects by extending a classical, competitive Moran model to incorporate an environment that fluctuates periodically in time. We adapt methods from work on these classical models to investigate the effects of the magnitude and frequency of environmental fluctuations on two important population statistics: the probability of fixation and the mean time to fixation.  In particular, we find that for small frequencies, the system behaves similar to a system with a constant fitness difference between the two species, and for large frequencies, the system behaves similar to a neutrally competitive model. Most interestingly, the system exhibits nontrivial behavior for intermediate frequencies. We conclude by showing that our results agree quite well with recent theoretical work on competitive models with a stochastically changing environment, and discuss how the methods we develop ease the mathematical analysis required to study such models.

%\gy{\begin{itemize}
%\item probably want to emphasize that our methods ease analysis
%\item maybe add a line stating that we specifically adapt the BKE from the constant-fitness-difference models
%\item should probably better motivate and define ``neutral model''
%\end{itemize}}

\section{Introduction}

The neutral theory of species diversity provides an important framework that facilitates the study of competition between species that are identically fit \cite{Hubbell1997,Hubbell2001,Hubbell2005,Hubbell2006}. Classic theoretical work has allowed explicit quantification of important ecological statistics, most notably the probability that a single species fixates, the average time it takes for a species to fixate, and species' average lifetimes \cite{Kimura1969,Crow1970,Ewens2012}. However, these results prompted criticism of neutral theory, as some species' lifetimes are estimated to be unrealistic (e.g., longer than the age of the earth \cite{Nee2005}).

More recently, and as a direct consequence of the criticisms of the results stemming from neutral theory, emphasis has been given to neutral models that incorporate fluctuating selective pressure due to a changing environment \cite{Chesson1981,Ashcroft2014,Danino2016,Danino2018,Danino2018b,Hidalgo2017,Kalyuzhny2015,Kessler2015,Wienand2017,Uecker2011}. These models generally assume that the population-level fitness of each species changes (positively or negatively) due to changes in the environment. While the specific mechanisms by which environmental fluctuations are incorporated into mathematical models varies from study to study, common assumptions are that the environment stochastically switches between a finite number of states, that each species' fitness is defined by the environmental state, and that no species has the highest fitness across all environmental states. The last assumption is in essence the so-called storage effect, which purports that no single species can be the most fit under all conditions in a changing environment \cite{Warner1985,Danino2018}. 

Importantly, nearly all theoretical work on neutral models in a fluctuating environment assumes that the environment changes stochastically, typically referred to as \emph{environmental stochasticity}. Here, we propose a deterministic alternative to environmental stochasticity that allows us to use standard theory to study the effects of environmental fluctuations as a function of model parameters. In particular, we consider a two-species Moran death-birth process that is parameterized by the difference in fitness between the two competing species, and we assume this fitness difference fluctuates sinusoidally over time. The process is consequently nonautonomous, and we show that established analytical methods from the autonomous neutral theory can be readily adapted to this time-periodic case. We therefore provide a simplified analytical framework through which the impact of environmental fluctuations on ecological competition can be studied. %\imp{emphasize that by taking a step back from stochastic to deterministic fluctuations, we can take advantage of normal PDE theory, which helps elucidate the mechanisms behind observed behaviors (e.g., large $\omega$ converges to neutral case)}}

Several motivations can be given for considering deterministic as opposed to stochastic fluctuations in this context. Many physical environmental variables, such as long-term climate fluctuations (e.g. ice age cycles), are essentially deterministic periodic fluctuations. Furthermore, the choice of deterministic fluctuations makes the problem much more mathematically tractable, in the sense that we may use the mean field theory developed in this work to predict and simulate much more easily. Finally, there is a well-known theory \cite{pavliotis2008multiscale,homogenization-papanicolaou-varadhan} connecting stochastic and deterministic fluctuations for elliptic and parabolic equations in the mathematics literature, which suggests that the deterministic results we consider here can be helpful in making predictions about situations with environmental stochasticity (especially in high-frequency limits).

The remainder of the paper is outlined as follows. In the following section, we review the classic neutral model as a Moran death-birth process, making use of the associated BKE to find the probability of fixation and average time to fixation. We then incorporate deterministic environmental fluctuations into the fitness terms of the model, resulting in a time-dependent BKE. Using intuition gained from the autonomous case, we argue that the system should asymptotically approach a periodic attractor, which quantifies the probability of fixation. We then use a similar argument to derive a method for determining the mean time to fixation. We validate our predictions by presenting numerical solutions of the BKE along with stochastic averages and show that they agree quite well. We then end our study by comparing our results with those from a neutral model with an analogous form of environmental stochasticity, and show that the two models qualitatively agree.

\section{Model}

\subsection{Moran process}
We consider a classic death-birth Moran process that describes the competition between two species in a population of constant size $N$ \cite{Chesson1981,Hubbell2005,Hubbell2006,Wright1931}. We denote by $N_1$ and $N_2=N-N_1$ the two species' respective sizes and by $f_1(s)$ and $f_2(s)$ their respective fitnesses at time $s$. Because our focus will be on competition with nonconstant fitnesses, we include explicit dependence on time in both species' fitness. Note that because $N_2$ is expressed as a function of $N_1$, this is a one-dimensional stochastic process. At each time step, two events occur: first, a single individual is chosen at random to die, then a single individual among the remaining individuals is chosen with probability proportional to each species' fitness to reproduce. This process can be expressed as a discrete-time death-birth Moran process with transition probabilities $P_{N_1}^+:=P(N_1+1|N_1)$ and $P_{N_1}^-:=P(N_1-1|N_1)$ defined as follows

\begin{equation*}
\begin{aligned}
P_{N_1}^+&=\frac{N_2}{N}\frac{f_1(s) N_1}{f_1(s)N_1+f_2(s) (N_2-1)},\\
P_{N_1}^-&=\frac{N_1}{N}\frac{f_2(s) N_2}{f_1(s) (N_1-1)+f_2(s) N_2},
\end{aligned}
\end{equation*}
and $P(N_1|N_1)=1-P_{N_1}^+-P_{N_1}^-.$

It is common practice and mathematically beneficial to rewrite these transition probabilities in terms of the proportion of the first species within the population $y(s)=N_1/N$ (and consequently $1-y(s)=N_2/N$). Writing $P_{y}^+:=P(y+1/N|y)$ and $P_{y}^-:=P(y-1/N|y)$, the transition probabilities become 

\begin{equation}\label{eq:Moran}
\begin{aligned}
P_{y}^+&=(1-y)\frac{f_1(s) y}{f_1(s) y+f_2(s) (1-y-1/N)},\\
P_{y}^-&=y\frac{f_2(s) (1-y)}{f_1(s) (y-1/N)+f_2(s) (1-y)},
\end{aligned}
\end{equation}
and $P(y|y)=1-P_{y}^+-P_{y}^-$. We remark here that if $f_1=f_2$, the stochastic process \eqref{eq:Moran} is called a \emph{neutral model} \cite{Hubbell1997}.

Model \eqref{eq:Moran} is an absorbing Markov process with two absorbing states, namely $y=0$ and $y=1$. If the system reaches $y=0$, we say population 1 is \emph{extinct}, and if the system reaches $y=1$, we say population 1 \emph{fixates}. Quantifying the probability of fixation is often a central aim of the analysis of such stochastic models \cite{Ewens2012}.  As such, we briefly review classic results on the fixation probably in model \eqref{eq:Moran} when the fitnesses $f_1$ and $f_2$ are constant, then dedicate the remainder of the paper to adapting these methods to determine the fixation probability and the mean time to fixation in model \eqref{eq:Moran} when $f_1$ and $f_2$ vary periodically in time.

\subsection{Constant fitness difference}\label{sec:const}

A central mathematical object in the study of both fixation probabilities and mean times to fixation in stochastic models with absorbing states is the backward Kolmogorov equation (BKE). The BKE associated with the Moran process \eqref{eq:Moran} is

\begin{equation}\label{eq:BKE}
\begin{aligned}
%-\frac{\partial u}{\partial t}&=-hx(1-x)\frac{a\cos{(2\pi\omega t+\phi)}}{1+a\cos{(2\pi\omega t+\phi)}(1-x)}\frac{\partial u}{\partial x} + \frac{h^2}{2}x(1-x)\frac{2+a\cos{(2\pi\omega t+\phi)}}{1+a\cos{(2\pi\omega t+\phi)}(1-x)}\frac{\partial^2 u}{\partial x^2}\\
-\frac{\partial u}{\partial t}&=x(1-x)\frac{f_1(Nt)-f_2(Nt)}{f_1(Nt)x+f_2(Nt)(1-x)}\frac{\partial u}{\partial x} + \frac{1}{2N}x(1-x)\frac{f_1(Nt)+f_2(Nt)}{f_1(Nt)x+f_2(Nt)(1-x)}\frac{\partial^2 u}{\partial x^2}\\
   &(x,t) \in  [0,1] \times (-\infty,\tau),\\
  &u(y,\tau\, |\, x,t) = \chi_1(y),
\end{aligned}
\end{equation} 
where $\chi_1(y)$ is the characteristic function of the set $\{1\}$ (see, e.g., \cite{Ewens2012} for a derivation). The temporal variable $t=s/N$ represents time on a generational scale, where a generation is assumed to be $N$ time steps in model \eqref{eq:Moran}, and allows us to consider time as approximately continuous for $N$ large \cite{Ewens2012,Wakeley2009}. Solutions $u(y,\tau\, |\, x,t)$ of this equation quantify the probability of the system entering the state $y=1$ by time $\tau$ assuming the system started in state $x$ at time $t$ \cite{Ewens2012,Risken1996}.

\begin{figure}[H]
{\centering
\includegraphics[width=0.5\textwidth]{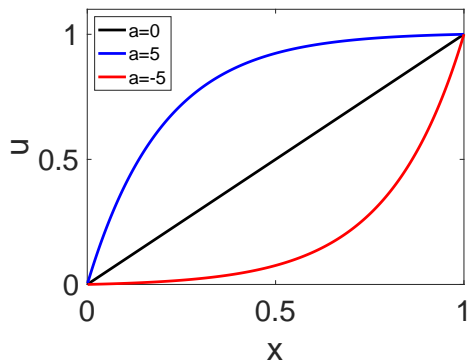}

}
\caption{The probability of fixation as a function of initial proportion $x$ for various fitness differences. Populations 1 and 2 have fitness $f_1=1+a/N$ and $f_2=1$, respectively. The figure shows the probability of fixation for $a=0$ (black), $a=5$, and $a=-5$.}
\label{fig:constantFixProbs}
\end{figure}

In the constant-fitness case, we will assume that the fitnesses can be written $f_1=1+a/N$ and $f_2=1$ so that the fitness difference between the two species is quantified by $a$. We scale the difference by a factor of $1/N$ in order to balance the drift and diffusion terms in the corresponding BKE, which we discuss below. In particular, if  $N\gg 1$, the BKE can be approximated up to order $1/N$ by 

\begin{equation*}
\begin{aligned}
-\frac{\partial u}{\partial t}&=\frac{a}{N}x(1-x)\frac{\partial u}{\partial x} + \frac{1}{N}x(1-x)\frac{\partial^2 u}{\partial x^2}\\
   &(x,t) \in [0,1] \times (-\infty,\tau),\\
  &u(y,\tau\, |\, x,t) = \chi_1(y).
\end{aligned}
\end{equation*}
Note that the entire right hand side is of order $1/N$ because of the choice to scale the fitness difference by $1/N$. We are interested in the value of $u(x,0)$ as we take $\tau \to \infty$; that is, the probability that a trajectory with initial proportion $x$ at time $t=0$ will ever reach the state $y=1$. This probability of fixation is given by the solution of the following steady state equation \cite{Ewens2012}
\begin{equation*}
\begin{aligned}
0&=\frac{a}{N}x(1-x)\frac{\partial u}{\partial x} + \frac{1}{N}x(1-x)\frac{\partial^2 u}{\partial x^2}\\
  &u(0)=0, \, u(1)=1,
\end{aligned}
\end{equation*}
which can be found explicitly to be
\begin{equation}
u(x)=
\begin{cases}
x & \text{if $a=0$}\\
\frac{1-e^{-ax}}{1-e^{-a}} &   \text{if $a\not=0$}. \\
\end{cases}
\end{equation}
In other words, under neutral competition ($a=0$), the probability that population 1 fixates is its initial proportion $x$ (Figure \ref{fig:constantFixProbs}, black curve). If $a>0$, population 1 has the fitness advantage and the probability of fixation is uniformly higher than the line $u(x)=x$ (Figure \ref{fig:constantFixProbs}, red curve). If $a<0$, population 2 has the advantage and the probability of fixation is uniformly lower than the same line (Figure \ref{fig:constantFixProbs}, blue curve).

\begin{figure}[H]
{\centering
\includegraphics[width=0.5\textwidth]{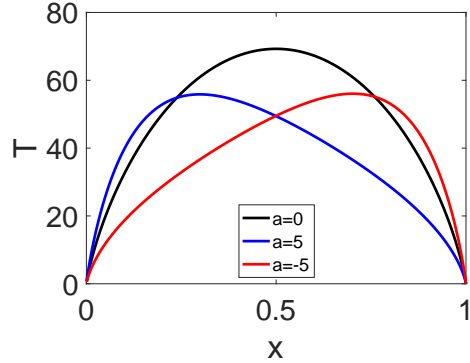}

}
\caption{The mean time to fixation as a function of initial proportion $x$.  for fixed fitness. Populations 1 and 2 have fitness $f_1=1+a/N$ and $f_2=1$, respectively. The figure shows the probability of fixation for $a=0$ (black), $a=5$, and $a=-5$.}
\label{fig:constantFixTimes}
\end{figure}

We can similarly determine the mean time to fixation; that is, the average time it takes for either species to fixate. If we denote the mean time to fixation by $T$, then $T$ is the solution to the boundary value problem \cite{Ewens2012}

\begin{equation*}
\begin{aligned}
-1&=\frac{a}{N}x(1-x)\frac{\partial T}{\partial x} + \frac{1}{N}x(1-x)\frac{\partial^2 T}{\partial x^2}\\
  &u(0)=0, \, u(1)=0.
\end{aligned}
\end{equation*}
 Numerically generated example solutions for $a=0$ (black), $a=5$ (blue), and $a=-5$ (red) with $N=100$ are shown in Figure \ref{fig:constantFixTimes} as functions of the initial proportion $x$. 

These results serve as a foundation upon which we base our analysis of system \eqref{eq:Moran} with time-dependent fitnesses. In the following sections, we adapt the methods above to study the fixation probabilities and mean times to fixation when the fitnesses vary periodically in time.

\subsection{Periodically fluctuating fitness}

The goal of this work is to incorporate the effects of a periodically fluctuating environment into a competition model in such a way that allows for efficient exploration of the effects of key parameters. As such, we will assume throughout the remainder of this paper that 
\begin{equation}\label{eq:fitnesses}
\begin{aligned}
f_1(s)&=1+\frac{a}{N}\cos{\left(2\pi\frac{\omega}{N^2} s+\phi\right)}\\
 f_2(s)&=1;
 \end{aligned}
 \end{equation} that is, one population's fitness oscillates around the other's fixed fitness, while preserving neutrality over long-term averages. The fitness difference between the two populations is now defined by the three parameters $a$, $\omega$, and $\phi$, which describe the magnitude, frequency, and initial phase shift of these fitness fluctuations, respectively. The phase shift $\phi$ determines which population has the initial fitness advantage. We scale the amplitude by a factor of $1/N$ for the same reason as in Section \ref{sec:const}. We scale the frequency by $1/N^2$ because doing so centers the nontrivial effects of the frequency around $\mathcal{O}(1)$, which we discuss with our results in Section \ref{sec:results}. We note that the model reduces to the neutral case if $a=0$, and to the constant fitness difference case if $\omega=0$. 
 
 Following the methods used in the constant-fitness case, we begin by considering the BKE \eqref{eq:BKE} with fitnesses given by \eqref{eq:fitnesses} up to $\mathcal{O}(1/N)$:
 \begin{equation}\label{eq:BKEh}
 \begin{aligned}
-\frac{\partial u}{\partial t}&=\frac{1}{N}x(1-x) a\cos(2\pi\omega t/N+\phi)\frac{\partial u}{\partial x}+ \frac{1}{N}x(1-x)\frac{\partial^2 u}{\partial x^2},\\
   &(x,t) \in [0,1]\times (-\infty,\tau),\\
  &u(y,\tau\, |\, x,t) = \chi_1(y).
  \end{aligned}
\end{equation} We note immediately that the BKE is no longer time-independent, and therefore we can no longer expect steady states to exist or be meaningful. The remainder of this paper will focus on the analysis of equation \eqref{eq:BKEh}.

\section{Results}\label{sec:results}

\subsection{Probability of fixation}\label{sec:fixProb}
Solutions of this equation \eqref{eq:BKEh} determine the probability that population $1$ fixates by time $\tau$ given that it had concentration $x$ at time $t$. As with the classical constant-fitness case, we are interested in the value of $u(x,0)$ as we take $\tau \to \infty$. In the classical neutral case, this corresponds to a time-invariant steady state solution of the BKE. However, the same argument is not valid when we consider time-dependent fitness, as we do not expect there to be any steady state of the equation with appropriate boundary values. Instead, we seek for a \emph{periodic attractor} for the system, namely we seek a solution of the equation

\begin{equation}\label{eqn:cell-problem}
\begin{aligned}
  -\frac{\partial u}{\partial t}&=\frac{1}{N}x(1-x) a\cos(2\pi\omega t/N+\phi)\frac{\partial u}{\partial x}+ \frac{1}{N}x(1-x)\frac{\partial^2 u}{\partial x^2}\\
  & (x,t) \in [0,1]\times \left(0,\frac{N}{\omega}\right),
\end{aligned}
\end{equation}
with boundary conditions
\begin{equation}\label{eq:perBCs}
\begin{aligned}
  u(0,t) &= 0, \, u(1,t) = 1,\\
  u(y,0) &= u\left(y,\frac{N}{\omega}\right),\, \text{ for all } (y,t)\in  [0,1]\times \left(0,\frac{N}{\omega}\right).
\end{aligned}
\end{equation}
This is the appropriate analog of the time-independent steady state in the case where we permit periodic forcing. This is because as $\tau \to \infty$ we anticipate that the solution to the backwards equation \eqref{eq:BKEh} will converge towards a function with the same period as the forcing term, which we impose with the periodic boundary condition in \eqref{eq:perBCs}.

\begin{figure}[H]
{\centering
\includegraphics[width=1\textwidth]{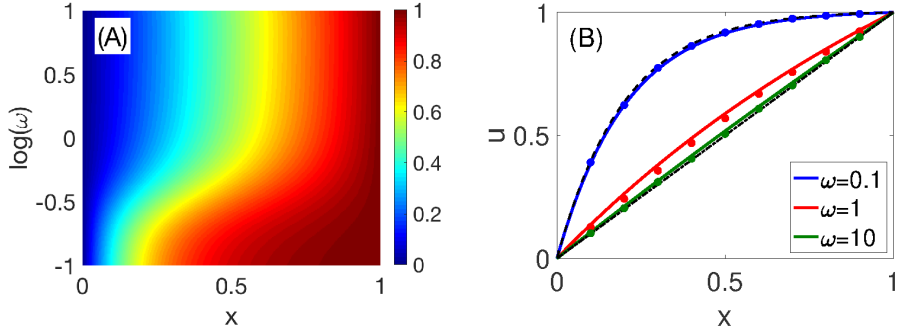}

}
\caption{The probability of fixation over $x$. In each figure, $a=5$ and $\phi=0$. A. The probability of fixation over varied initial proportion $x$ and environmental fluctuation frequency $\omega$ on a logarithmic scale.  B. The probability of fixation over initial proportion $x$ for $\omega=0.1$ (blue), $\omega=1$ (red),  and $\omega=10$ (green).  The curves are generated by the PDE system \eqref{eqn:cell-problem}; the circles correspond to the probability of fixation averaged over 20000 stochastic simulations with $N=100$. The black dashed curve is the probability of fixation from the constant fitness difference case shown in Figure \ref{fig:constantFixProbs}, the black dash-dotted curve is the probability of fixation in the neutral case.}
\label{fig:fixProbs}
\end{figure}

We do not anticipate that the solution of equation \eqref{eqn:cell-problem} will have a simple closed-form. However, one can solve this problem using standard finite element software. Example solutions computed using FEniCS \cite{Fenics1,Fenics2} are shown in Figure \ref{fig:fixProbs}. We remark that this equation is degenerate, and rigorous numerics are quite challenging \cite{Duan2019}, but in practice the standard solvers seem to be sufficient.

We calculate the probability of fixation over three important quantities: the initial proportion $x$, the environmental fluctuation frequency $\omega$, and the initial phase shift $\phi$. Figure \ref{fig:fixProbs}A shows an example solution of PDE \eqref{eqn:cell-problem} over $x$ and $\log(\omega)$ with $a=5$, $N=100$, and $\phi=0$  fixed. This choice of $\phi$ gives population 1 the early-time advantage. Of course, the probability of fixation is always $0$ when $x=0$. For small $\omega$, the fixation probability quickly increases with $x$. As $\omega$ increases, the rate at which the fixation probability increases with $x$ slows. Example fixation probability curves for various fixed values of $\omega$ are shown in Figure \ref{fig:fixProbs}B. For $\omega=0.1$ (Figure \ref{fig:fixProbs}B, blue curve), the environmental fluctuations are slow, and population 1's initial advantage persists for a relatively long time, and the probability of fixation is large relative to larger $\omega$ values. Unsurprisingly, the probability of fixation for $\omega$ small is close to the probability of fixation in the case where population 1 has a constant fitness advantage (black dashed line). This is because the environmental fluctuations happen on a timescale that is longer than the time it takes for either species to fixate. At the other extreme, the probability of fixation tends to that of the neutral case for $\omega$ large. This is illustrated by the green curve in Figure \ref{fig:fixProbs}B, corresponding to the probability of fixation when $\omega=10$, which lies nearly on top of the black dashed-dotted line that defines the probability of fixation in the neutral case. This suggests that as $\omega$ becomes large, the environmental fluctuations play no effect on the outcome of competition. Nontrivial behavior occurs for intermediate values of $\omega$. The red curve in Figure \ref{fig:fixProbs}B defines the probability of fixation when $\omega=1$ and lies between the $\omega=0.1$ and $\omega=10$ cases. This case does not have any apparent analog to competition without environmental fluctuations. For the sake of comparison, we also present the probability of fixation found by averaging over 20000 stochastic simulations with $N=100$, given by the colored circles along each curve in Figure \ref{fig:fixProbs}B. All stochastic simulations were computed using a standard Gillespie algorithm \cite{Gillespie1977,Erban2007}.

\begin{figure}[H]
{\centering
\includegraphics[width=1\textwidth]{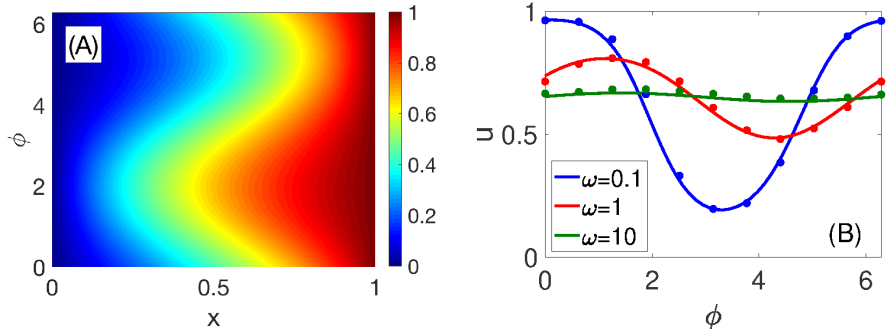}

}
\caption{The probability of fixation over $\phi$ with $a=5$. A. The probability of fixation over varied initial proportion $x$ and initial phase shift $\phi$, for $\omega=1$ fixed. B. The probability of fixation over initial phase shift $\phi$ for $\omega=0.1$ (blue), $\omega=1$ (red),  and $\omega=10$ (black), with $x=2/3$ %\gy{(NOTE: $x$ was 1/3, but the change in convention changed it to 2/3)}
 fixed. The curves are generated by the PDE system \eqref{eqn:cell-problem}; the circles correspond to the probability of fixation averaged over 20000 stochastic simulations with $N=100$.}
\label{fig:fixProbsPhi}
\end{figure}

Figure \ref{fig:fixProbsPhi} shows the probability of fixation over varied $\phi$. Figure \ref{fig:fixProbsPhi}A shows an example solution of PDE \eqref{eqn:cell-problem} over $x$ and $\phi$ with $a=5$, $N=100$, and $\omega=1$ fixed. In Figure \ref{fig:fixProbsPhi}B, we show the probability of fixation across varied phase shift $\phi$ with $x=2/3$ for three fixed frequencies $\omega$. As with the probabilities of fixation in Figure \ref{fig:fixProbs}, the probability of fixation seems to converge to the neutral case as $\omega$ gets large: for $\omega=10$, $u\approx2/3$ over all $\phi$, which is the probability of fixation in the neutral case without fluctuations.

\subsection{Mean time to fixation}\label{sec:ttf}

One can use a similar method to compute mean time to fixation. We outline the derivation briefly here (cf. \cite{Lange2010}). Let $T_t$ be a random variable representing the time remaining until fixation from time $t$, and let $s>t$ be some future time. Then the mean time to fixation for the system starting at state $x$ at time $t$, $T(x,s)$, satisfies
\begin{equation} \label{eqn:Transition-MPT}
\begin{aligned}
  T(x,t) &:= \E[T_t | X_t = x] \\
  &\approx (s-t) + \E[T_s | X_t = x] \\
  &= (s-t) + \E[w(s,X_s) | X_t = x],
  \end{aligned}
\end{equation}
%\imp{Is this a new $u$?}
where the last equality follows from the law of total expectation and the ``$\approx$'' is due to the possibility that fixation occurred during the interval $[s,t]$ (an event with probability $o(s-t)$, which we can neglect in the limit).
Taking $s \to t$ and using Ito's rule then gives the equation
\begin{equation}
\begin{aligned}
  0 &= \partial_t T + \frac{1}{N}x(1-x) a \cos( 2\pi \omega t/N+\phi) T_x + \frac{1}{N}x(1-x) T_{xx} + 1.
   %&= \partial_t u + \frac{x(1-x)}{1 + (1-x)a \sin \omega t}\left( a \sin( \omega t+\phi) u_x + \frac{h(2 + a \sin \omega t)}{2} u_{xx} \right) + 1
  \end{aligned}
  \label{eqn:Backwards-MPT}
\end{equation}
In the case where we study mean time to fixation of either species, the appropriate boundary conditions are $u(0,t) = u(1,t) = 0$, namely a Dirichlet condition.

\begin{figure}[H]
{\centering
\includegraphics[width=1\textwidth]{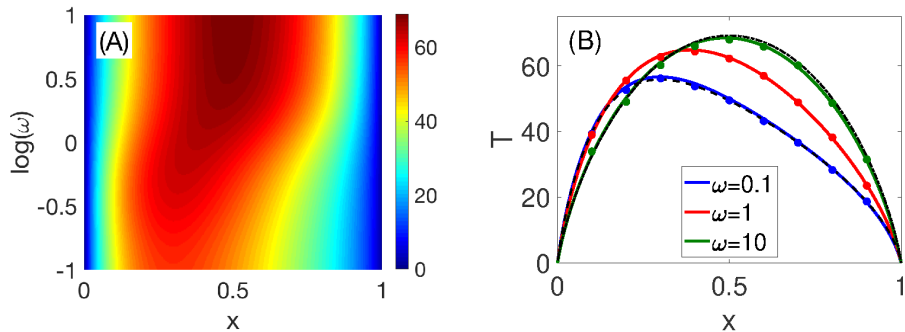}

}
\caption{The mean time to fixation over $x$. In each figure, $a=5$ and $\phi=0$. A. The mean time to fixation over varied initial proportion $x$ and environmental fluctuation frequency $\omega$ (log scale). B. The mean time to fixation over initial proportion $x$ for $\omega=0.1$ (blue), $\omega=1$ (red),  and $\omega=10$ (black). The curves are generated by the PDE system \eqref{eqn:Backwards-MPT}; the circles correspond to the probability of fixation averaged over 20000 stochastic simulations with $N=100$. The black dashed curve is the mean time to fixation in the constant fitness difference case shown in Figure \ref{fig:constantFixTimes}, the black dash-dotted curve is the mean time to fixation in the neutral case.}
\label{fig:ttf}
\end{figure}

We visualize a numerically generated solution of \eqref{eqn:Backwards-MPT} as a surface over varied $x$ and $\log(\omega)$ with $\phi=0$ fixed in Figure \ref{fig:ttf}A, with one-dimensional curves over $x$ for various fixed values of $\omega$ in \ref{fig:ttf}B. As with the probability of fixation (Figure \ref{fig:fixProbs}), the mean time to fixation for small $\omega$ closely agrees with the mean time to fixation for the system with a constant fitness difference. In particular, the blue curve in Figure \ref{fig:ttf}B represents the mean time to fixation when $\omega=0.1$, and matches the black dashed curve, which represents the mean time to fixation in the constant fitness difference case (Figure \ref{fig:constantFixTimes}). Similarly, the mean time to fixation once again tends to that of the neutral case when $\omega$ is large. The green curve shows the mean time to fixation when $\omega=10$, which agrees well with the black dash-dotted curve representing the mean time to fixation in the neutral case. This corroborates the conclusion from Section \ref{sec:fixProb} that for large $\omega$, environmental fluctuations do not affect the outcome of competition. Between these two extremes, we again observe nontrivial behavior. The red curve represents the mean time to fixation when $\omega=1$, and does not correspond to any fixed-fitness-difference case.

\begin{figure}[H]
{\centering
\includegraphics[width=1\textwidth]{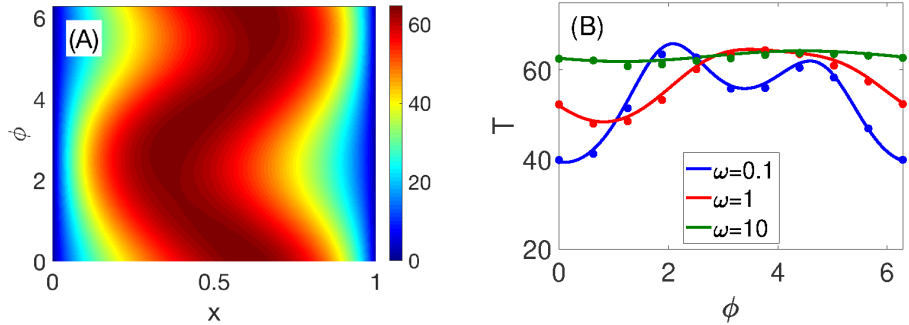}

}
\caption{The mean time to fixation over $\phi$. In each figure, $a=5$. A. The mean time to fixation over varied initial proportion $x$ and initial phase shift $\phi$, for $\omega=1$ fixed. B. The mean time to fixation over initial phase shift $\phi$ for $\omega=0.1$ (blue), $\omega=1$ (red),  and $\omega=10$ (black), with $x=2/3$ fixed. The curves are generated by the PDE system \eqref{eqn:Backwards-MPT}; the circles correspond to the probability of fixation averaged over 20000 stochastic simulations with $N=100$.}
\label{fig:ttfPhi}
\end{figure}

%The figure was generated by solving the PDE
%
%\begin{equation}
%  -u_t = \frac{1}{N} a x(1-x) \cos(2\pi \omega t/N+\phi)u_x + \frac{1}{N}x(1-x) u_{xx}
%  \label{eqn:PDE-mean-fix}
%\end{equation}
%for $a = 5$ and $\omega = 1$. Specifically, this graphic is obtained by running the PDE backwards in time for some number of periods until it converges to a periodic attractor. The last computed period is then displayed (meaning the $y$ axis represents time rescaled by $\omega$).

\subsection{Comparison to environmental stochasticity}

Incorporating the effects of a changing environment using deterministic periodic fitness fluctuations as above allows for simplified analysis when compared to stochastic fluctuations. Of course, there are many environmental changes that are modeled stochastically \cite{Chesson1981,Ashcroft2014,Danino2016,Danino2018,Danino2018b,Hidalgo2017,Kalyuzhny2015,Kessler2015,Wienand2017}. We therefore present a comparison of our results with those from an analogous competition model with a stochastically varying environment, and show that the qualitative behaviors of the two models match. We emphasize that our methods developed above provide a more tractable approach while capturing many of the relevant behaviors of models with a stochastically varying environment.

\begin{figure}[H]
{\centering
\includegraphics[width=1\textwidth]{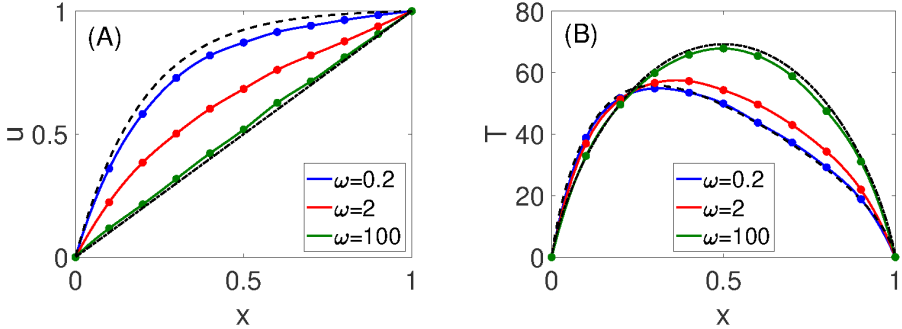}

}
%\caption{Probability of fixation and mean time to fixation with environmental stochasticity. In both figures, $a_0=5.$ Each colored circle represents the quantity (probability of fixation or mean time to fixation) averaged over 20000 simulations, and each curve is generated by interpolating these points. A. The probability of fixation over the initial proportion $x$ for various values of $\omega$. Each colored circle represents the probability of fixation averaged over 20000 simulations, and each curve is generated by interpolating these points. For $\omega=0.2$ (blue), the probability of fixation closely matches that of the constant fitness difference case with $a=5$ (black dashed curve). For $\omega=100$ (green), the probability of fixation agrees well with that of the neutral case (black dash-dotted curve). For the intermediate value of $\omega=2$ (red), the probability of fixation lies uniformly in between the blue and green curves.  B. The mean time to fixation over the initial proportion $x$ for various values of $\omega$. The curves corresponding to $\omega=0.2$ (blue) and $\omega=100$ (green) once again agree well with the mean time to fixation in the fixed-fitness-difference case (black dashed curve) and neutral case (black dash-dotted curve), respectively. The curve corresponding to $\omega=2$ again displays a nontrivial behavior. \gy{trim this caption down...}}
\caption{Probability of fixation (A) and mean time to fixation (B) over initial proportion $x$ for various environmental stochasticity frequencies $\omega$. In both figures, $a_0=5$ and $a(0)=a_0$. Each colored circle represents the quantity averaged over 20000 simulations, and each curve is generated by interpolating these points.  For small frequencies ($\omega=0.2$, blue curves), the probability of fixation and mean time to fixation closely match that of the constant fitness difference case with $a=5$ (black dashed curves). For large frequencies ($\omega=100$, green curves), the probability of fixation and mean time to fixation agree well with that of the neutral case (black dash-dotted curves). For intermediate frequencies ($\omega=2$, red curves), the probability of fixation and mean time to fixation lie between the blue and green curves, displaying nontrivial behavior.}
%The curves corresponding to $\omega=0.2$ (blue) and $\omega=100$ (green) once again agree well with the mean time to fixation in the fixed-fitness-difference case (black dashed curve) and neutral case (black dash-dotted curve), respectively. The curve corresponding to $\omega=2$ again displays a nontrivial behavior. \gy{trim this caption down...}}
\label{fig:envStoch}
\end{figure}

We again consider a Moran process of the form \eqref{eq:Moran}, with $f_1=1+a(t)/N$, $f_2=1$, where $a(t)$ varies stochastically between two constant states, say $a_0$ and $-a_0$, at probability rate $\omega/N^2$. This is a natural analog of the continuous fluctuations considered throughout this work. As above, we again consider the effect of the frequency of environmental changes on the probability of fixation and the mean time to fixation, visualized in Figure \ref{fig:envStoch}. Figure \ref{fig:envStoch}A shows the probability of fixation for small, intermediate, and large frequencies. Each colored circle represents the probability of fixation averaged over 20000 simulations, and each curve is generated by interpolating these points. These curves agree well with the continuous-fluctuations case: for small $\omega$ (blue curve), the probability of fixation is close to the fixed-fitness-difference case (black dashed curve). For large $\omega$ (green curve), the probability of fixation is almost exactly that of the neutral case (black dash-dotted line). Nontrivial behavior is again observed for intermediate $\omega$ (red curve). Similar qualitative agreement in the mean time to fixation between the continuous and stochastic environmental fluctuations can be seen in Figure \ref{fig:envStoch}B. Each colored circle corresponds to the mean time to fixation averaged over 20000 simulations, and each curve is again generated by interpolating these points. Again, for small $\omega$ (blue curve), the mean time to fixation is agrees well with the case in which the fitness difference is constant (black dashed curve), while for large $\omega$, the mean time to fixation is agrees well with the neutral case (black dashed-dotted line). Intermediate values of $\omega$ once again produce behavior that does not correspond to any fixed-fitness case (red curve).

The behavior of the system with environmental stochasticity overall agrees well with the behavior of the system with deterministic fluctuations.  This suggests that, in some settings, studying a system with a deterministically fluctuating environment might serve as a reasonable and much more tractable proxy for environmental stochasticity.  In particular, we remark here that our methods provide a numerically efficient way of studying these types of systems compared to alternative methods. In particular, numerically solving the PDEs \eqref{eqn:cell-problem} and \eqref{eqn:Backwards-MPT} is much faster computationally than measuring the averages by iterating the associated Moran process with or without environmental stochasticity.

\section{Discussion}

We have investigated the effects of a continuously fluctuating environment on two neutrally-competitive species by allowing each species' reproductive fitness to vary periodically over time, while maintaining the same long-term average. Our framework can be considered a simplification of recent and classic methods for analyzing competition within a stochastically variation environment \cite{Chesson1981,Ashcroft2014,Danino2016,Danino2018,Danino2018b,Hidalgo2017,Kalyuzhny2015,Kessler2015,Wienand2017}. By considering continuous-time environmental variation, we are better able to adapt and utilize standard analytical approaches to calculating the probability of fixation and mean time to fixation of either species in the competitive process. We validate our analytical results by showing that numerical averages from stochastic simulations match both the fixation probability and mean time to fixation. Moreover, we show that the qualitative behavior of the system with continuous-time environmental fluctuations agrees well with that of the system that stochastically switches between two environmental states. 

Our results confirm the intuitively clear principle that if the environment changes on a timescale much slower than that of the population dynamics, the population with the early-time fitness advantage should outcompete the other population with the same probability as a population with a fixed fitness advantage; that is, a fitness advantage independent of environmental variation.  Moreover, we find that if an environment fluctuates very rapidly, the resulting effects on the population dynamics homogenize, and the system behaves identically to the classic neutral case, in which both populations' fitnesses are equal and independent of time. Intermediate frequencies provide a new regime, unique from any constant-fitness case. When the environmental fluctuations occur on these intermediate timescales, the probability of fixation and mean time to fixation both depend not only on initial proportion $x$ and the initial fitness difference, but on the frequency of environmental fluctuations.

Our analysis is centered on the backward Kolmogorov equation (BKE). In time-independent stochastic processes, it is well known the probability of fixation and the mean time to fixation are quantified by certain steady states of the BKE \cite{Ewens2012,Risken1996}. Of course, incorporating environmental fluctuations necessarily results in a time-dependent process, and classic methods for determining these two statistics become ineffective. However, we have developed here a straightforward modification of these existing methods that yields the analogous results when the stochastic process is periodic in time.

Similar methods were presented in  \cite{Uecker2011} to study \emph{beneficial} mutations in changing environments. The authors adapted a method using generating functions to find an explicit approximation to the probability of fixation in a variety of temporally-changing environments. Their approximation yields very nice results for competition in which one population (the mutant) has a long-term fitness advantage, but the authors comment that their approximation yields a fixation probability of zero if the long-term average of both species' fitness is equal, which ``underestimates the true value''. Our present work captures this outstanding case, and further agrees with Uecker et.\ al's finding that for high-frequency environmental oscillations, the system behaves very similarly to the corresponding system without environmental variation.

This work can function as a bridge between two active areas of study within theoretical ecology: competition within a variable environment and stochastic competition with demographic fluctuations. Most work that considers competition in a changing environment assumes the total population of all competitors is constant \cite{Chesson1981,Ashcroft2014,Danino2016,Danino2018,Danino2018b,Hidalgo2017,Kalyuzhny2015,Kessler2015,Wienand2017}. This assumption greatly simplifies the resulting mathematics by reducing the number of stochastic variables by one. Separately, there is a growing literature on stochastic competition models that relax this constant-population assumption, but do not consider environmental effects \cite{Lin2012,Constable2015,Huang2015,Chotibut2017,Constable2016,Constable2018,Czuppon2018,Czuppon2018b,Young2018}. A natural way to simultaneously approach both a fluctuating environment and a variable population size is to incorporate the changing environment deterministically, as we do in this work.

%\section{Appendix}
%
%\subsection{BKE derivation}
%Let $u(y,\tau|x,t)$ denote the probability that the system is in state $y$ at time $\tau$ given that it was in state $x$ and time $t$. The master equation corresponding to the stochastic process \eqref{eq:Moran} can be written
%
%
%\begin{equation}\label{eq:master}
%-\frac{\partial u(y,\tau\, |\, x,t)}{\partial t}=P^{+}_{x}u(y,\tau\, |\, x+1/N,t)+P^{-}_{x} u(y,\tau\, |\, x-1/N,t)-\left(P^{+}_{x}+P^{-}_{x} \right)u(y,\tau\, |\, x).
%\end{equation}
%
%Expanding equation \eqref{eq:master} in powers of $1/N$ up to order $\mathcal{O}(1/N^2)$, we recover the backward Kolmogorov equation (BKE)
%
%\begin{equation}\label{eq:BKEapp}
%\begin{aligned}
%-\frac{\partial u}{\partial t}&=\frac{1}{N}\left[P_x^+(x)-P_x^-(x)\right]\frac{\partial u}{\partial x}+ \frac{1}{2N^2}\left[P_x^+(x)+P_x^-(x)\right]\frac{\partial^2 u}{\partial x^2}\\
%&=\frac{1}{N}x(1-x)\frac{f_1-f_2}{f_1x+f_2(1-x)}\frac{\partial u}{\partial x} + \frac{1}{2N^2}x(1-x)\frac{f_1+f_2}{f_1x+f_2(1-x)}\frac{\partial^2 u}{\partial x^2}\\
%&=\frac{1}{N}x(1-x)\frac{a\cos{(2\pi\omega t+\phi)}}{1+a\cos{(2\pi\omega t+\phi)}x}\frac{\partial u}{\partial x} + \frac{1}{2N^2}x(1-x)\frac{2+a\cos{(2\pi\omega t+\phi)}}{1+a\cos{(2\pi\omega t+\phi)}x}\frac{\partial^2 u}{\partial x^2}.
%\end{aligned}
%\end{equation} 

\bibliographystyle{unsrt}
\bibliography{WFbib}

\end{document}